# Gene & Genome Duplication in *Acanthamoeba Polyphaga Mimivirus*

Karsten Suhre

19/07/2005

Information Génomique & Structurale, UPR CNRS 2589, 31, chemin Joseph-Aiguier, 13402 Marseille Cedex 20, France, e-mail: karsten.suhre@igs.cnrs-mrs.fr

## Abstract

Gene duplication is key to molecular evolution in all three domains of life and may be the first step in the emergence of new gene function. It is a well recognized feature in large DNA viruses, but has not been studied extensively in the largest known virus to date, the recently discovered *Acanthamoeba Polyphaga Mimivirus*. Here we present a systematic analysis of gene and genome duplication events in the Mimivirus genome. We find that one third of the Mimivirus genes are related to at least one other gene in the Mimivirus genome, either through a large segmental genome duplication event that occurred in the more remote past, either through more recent gene duplication events, which often occur in tandem. This shows that gene and genome duplication played a major role in shaping the Mimivirus genome. Using multiple alignments together with remote homology detection methods based on Hidden Markov Model comparison, we assign putative functions to some of the paralogous gene families. We suggest that a large part of the duplicated Mimivirus gene families are likely to interfere with important host cell processes, such as transcription control, protein degradation, and cell regulatory processes. Our findings support the view that large DNA viruses are complex evolving organisms, possibly deeply rooted within the tree of life, and oppose the paradigm that viral evolution is dominated by lateral gene acquisition, at least in what concerns large DNA viruses.

**Introduction**

It has long been realized that new gene material frequently emerges through gene and genome duplication (26, 27). The precise mechanisms of these events are diverse, each leaving its own particular signature in the genome (for a recent review see 36). Once a gene has been duplicated it may be subject to basically three different types of fate: nonfunctionalization, where one of the two copies of a duplicate pair degenerates into a pseudogene and may subsequently be lost from the genome (18, 19), subfunctionalization, which consists in the division of the original functions of the ancestral gene between the two duplicates (9), and neofunctionalization where one copy in a duplicate pair acquires a new function (37). Eventually, divergent evolution may lead to a point where homologies between two genes of common ancestry become difficult or impossible to detect (11). The unexpectedly small structural variation between different protein families that has been unveiled by the recent large structural genomics efforts (31) corroborates this observation, suggesting that the prevalence of gene duplication in all three domains of life (36) is even larger than previously thought.

The recent discovery (16) and subsequent genome sequencing (29) of the largest known virus to date, *Acanthamoeba Polyphaga Mimivirus,* has raised a number of fundamental questions about what was thought so far as being established boundaries between viruses and cellular life forms (5, 7, 14). In particular, the size of the Mimivirus virion is comparable to that of a mycobacterium. Its genome, containing close to 1.2 million nucleotides and coding for 911 predicted proteins, holds more than twice as much genetic information than what suffices to small bacteria for life. Moreover, the Mimivirus genome hosts a wide spectrum of genes that have never been found in such combination in a virus, in particular a large set of genes related to protein transcription and translation. On the other hand, what is rather common for a viral genome is the fact that a large fraction of the Mimivirus genes displays only weak or no homology to any other known genes in the databases. For only one third (298 / 911) of the Mimivirus genes Raoult *et al.* (29) were able to assign putative functions, while this ratio is much higher for the genomes of all fully sequenced "living" organisms.

Here we set out to investigate the question as to how many of these genes of unknown origin may have been generated through duplication processes within the Mimivirus genome itself, and how these duplications may then have shaped the Mimivirus genome. The aim of this work is to identify and characterize events of gene and genome duplication in the Mimivirus genome in order to shed new light on the origin of Mimivirus' exceptionally large size and on the importance of gene duplication in large DNA viruses in general. In the following we report evidence for an ancient event of duplication of a large part of the Mimivirus chromosome, as well as for numerous tandem gene duplication events, and we will show that some of these duplication events may play a role in virus-host adaptation.

**Results**

**One third of the Mimivirus genes has at least one paralogue in the genome.**
We compared all 911 predicted Mimivirus genes against each other using the sequence alignment software BLAST (1) to identify genes that have significant matches in the genome. Search for paralogous genes was iterated until convergence using position specific weight matrices constructed from the set a homologous genes found in each previous step as implemented in the Psi-BLAST version of BLAST. 347 paralogous genes in 77 families were detected with this method when applying a conservative detection cutoff e-value in the (Psi-)BLAST search of $10^{-10}$. When using a more permissive ($10^{-5}$) or a more stringent ($10^{-25}$) e-value, 398 and 244 paralogous genes in 86 and 58 families, respectively, were detected. Thus, between 26.3% and 35.0% of the Mimivirus genes have at least one homologue in the virus' genome, depending on the choice of the e-value cut-off. To test for a possible dependence on gene annotation, the Mimivirus genome was split into non-overlapping segments of 1000 nucleotides in length. These segments were compared to segments of the same size, but overlapping by 50% with each other, using BLAST at the nucleotide level (BLASTN) and at the amino acid level after translation in all 6 reading frames (TBLASTX). The results were comparable to those found using BLAST at the gene level (BLASTP) in what concerns our conclusions with respect to the overall genome and gene duplications, except that these methods were less sensitive and yielded less hits at lower sequence identity levels, especially in the BLASTN case. As these computations did not reveal any unexpected new insight, but confirmed the robustness of the approach with respect to the applied detection algorithm (BLASTP), BLASTN and TBLASTX results will not be further presented in this paper.

**The orientation and location of gene duplication events is not random.**

The Mimivirus genome is coded on a linear chromosome, that may adopt a circular topology through non-covalent interactions between two 900 nt long repeated sequences near the chromosome ends, as observed in some other large DNA viruses (29). The fraction of duplicated genes that are inserted in parallel orientation to the coding direction of the matching gene (*cis*) is with 20.2% (22.1% for $e=10^{-5}$, 16.4% for $e=10^{-25}$) nearly twice as high as the fraction of genes that are duplicated in anti-parallel orientation (*trans*), which is 11.7% (12.9% for $e=10^{-5}$, 9.95%, for $e=10^{-25}$). 61% of all pairs of genes that are duplicated in *trans* are located on different halves of the Mimivirus chromosome, whereas 79% of the duplications in *cis* occur on the same chromosome half. A large number of tandem, or near-tandem, gene duplications were detected, the most striking case consisting in an eleven-fold duplication of genes L175 to L185 (dubbed *Lcluster* in the following; see Table 1 for gene location and orientation of the largest families of paralogues). The overall trend is that *cis* duplications are more localized (often tandem or near-tandem) while *trans* duplications are more likely to occur across the chromosome center. This trend becomes visible when connecting corresponding best-matching pairs that are duplicated in *cis* and *trans*, respectively (Figure 1).

**Evidence for a segmental duplication of a large telomeric chromosome fraction.**

Figure 2 shows a zoom into the "telomeric" regions of the Mimivirus chromosome. Remnants of chromosomal synteny can be identified between 5'-position 0 and 5'-position 110,000 and the corresponding 3'-position 110,000 to 3'-position 220,000 and also between 5'-120,000 to 5'-200,000 and 5'-0 to 5'-80,000. Overlapping with these is a synteny between the 5'-20,000 to 5'-110,000 and 3'-0 to 3'-100,000. The exact history of this (or these) segmental genome duplication events is difficult to reconstruct, as it is overlain by numerous local *cis*-duplication events and as no information is available on potential gene deletions in this context. One parsimonious explanation could be a segmental duplication of an about 200,000 nt long telomeric chromosome fraction, followed by a rearrangement (immediately or later) around its center. Interestingly three tRNA-Leu genes are found duplicated in concert with this (these) event(s). They are highly conserved (displaying only 4 point mutations), while the adjacent genome regions accumulated such a large number of mutations that homology at the nucleotide level becomes difficult to be identified. Figure 3 shows the frequency distribution of all gene duplication events. A pronounced maximum for *trans* duplications is observed at a sequence identity level of 25% that characterizes the segmental gene duplication as a more ancient event. *Cis* duplications also peak at this value and are likely to correspond to older tandem duplications events. A second pronounced maximum at the 50% sequence identity level for *cis* duplications suggests a more recent origin for the corresponding tandem duplications (i.e. the *Lcluster*).

**Duplicated genes can be used to detect remote homologies and to improve on the functional gene annotation**

In every genome sequencing project the question of how to annotate putative genes has to be addressed. It is standard procedure to compare all predicted genes to existing annotated databases (e.g. SWISS-PROT (4) or the non redundant protein database at NCBI) using sequence-to-sequence comparison tools, most often BLAST. More sensitive methods, that also allow the identification of more remote functional relationships, are based on sequence-to-profile comparison. These include tools like reverse position specific BLAST (rpsBLAST) (1) and hmmer (8), which compare a query gene to an annotated aligned gene family rather than to a single gene (see material and methods for a selection of generally used protein family databases). Depending on the quality of the resulting hits, manual quality checks and further refinement is done, usually based on multiple alignments and possibly phylogentic tree reconstructions, in order to verify the predicted orthologies to a gene or gene family of known function. The result of this procedure is what is commonly known as the "genbank annotation" of a genome. In the case of Mimivirus (and this is true for all virus genome sequencing projects), no function could be attributed convincingly to a large number of genes using this procedure. These genes are thus annotated as "hypothetical", supplemented in some cases by a description of a generic feature of that gene, such as specific type of repeat (ankyrin, triple helix collagen repeat, leucine rich repeat).

However, in the case where multiple copies of a gene are found in the genome, the idea of using profile or Hidden Markov Model (HMM) search methods can be taken a step further. Different such methods have recently been developed (30, 32, 35). They allow the comparison of an aligned set of genes (the paralogous genes) to a database of annotated profiles or HMMs with much higher sensitivity than sequence-to-sequence and sequence-to-profile comparisons. Here we use the HHsearch software (32), which, in addition to HMM-HMM comparison, evaluates the correspondence between the predicted secondary protein structure of the query protein and those of the potential hits (using observed structure information from the Protein Data Bank (PDB) where available). The result of a HHsearch for a single family of paralogues is then a list of hits, ranked by the probability that a hit is a true positive. For all families of paralogues, these results, together with the corresponding multiple alignments that were used to build the HMMs, are available as supplementary material at http://igs-server.cnrs-mrs.fr/suhre/mimiparalogues/ . This dataset may serve as a starting point for further analysis of a given Mimivirus paralogue family.

**Some of the larger paralogous families are related to virus-host interactions.**
Figure 4 (left) shows the position of all paralogous genes by their position on the chromosome. Hotspots of local tandem duplication activities can be detected and are particularly pronounced for the gene family N172 (*Lcluster*). A clustered view of all genes is given in Figure 4 (right) and the largest families are listed in Table 1. By far the largest paralogous gene family with 66 members contains the ankyrin double helix repeat proteins (N14). Ankyrin repeat containing proteins are ubiquitously found in large paralogous families in both, viral and bacterial genomes. These genes are thought to play structural roles in the cell and are not further discussed here.

The second largest family (N35) contains 26 genes that are all annotated as unknown (4 of them contain WD-repeats). However, using remote homology detection methods together with advanced multiple alignment techniques (see methods) we find that all of these proteins contain a common, about 170 amino acids long N-terminal domain that clearly matches the BTB/POZ domain. The BTB/POZ domain mediates homomeric dimerisation and in some instances heteromeric dimerisation. POZ domains from several zinc finger proteins have been shown to mediate transcriptional repression and to interact with components of histone deacetylase co-repressor complexes. Best matches to proteins with known structure are the promyelocytic leukaemia zinc finger protein (PDBid 1buo) and the B-cell lymphoma 6 protein (PDBid 1r28). The genes from the N35 paralogue family are thus likely to play a role in transcriptional regulation.

The third largest cluster (N172, *Lcluster*) is also the most exceptional in what concerns its eleven fold tandem repeat of proteins. A multiple alignment of these genes indicates that they code for real proteins and that these proteins are likely under selective pressure. For instance, amino acid type is often conserved within aligned columns, and stretches without any insertions and deletions are followed indel-rich regions (signatures of structures elements and loop regions, respectively). However, no clear function could be attributed to this cluster, and it has no significant match outside the Mimivirus genome. The highest scoring hits from remote homology detection, albeit well below certainty levels in what concerns the probability that these are true positives, are sometimes linked to interaction with RNA.

The cluster N165, that is found close to the Lcluster, also contains only genes that are annotated as unknown, most of them containing several Pfam FNIP repeats. Again, using remote homology detection, we can identify an N-terminal domain that matches the Pfam F-box domain, which is a receptor for ubiquitination targets. This relatively conserved structural motif is present in numerous proteins and serves as a link between a target protein and a ubiquitin-conjugating enzyme. The SCF complex (e.g., Skp1-Cullin-F-box) plays a similar role as an E3 ligase in the ubiquitin protein degradation pathway. Different F-box proteins as a part of SCF complex recruit particular substrates for ubiquitination through specific protein-protein interaction domains. Interestingly, several copies of ubiquitin-conjugating enzymes are also present in the Mimivirus genome (i.e. gene L460) as well as a ubiquitin-specific protease (R319). Thus, the genes in cluster N165 can be predicted to play a role in protein degradation using the ubiquitin pathway.

For cluster N226 little can be said at present. Cluster N232 on the other hand contains genes that are predicted as protein kinases and may thus play a role in different cell regulatory processes.
Other notable families, not discussed in more detail here, are family N137, which contains proteins with glycosyl-transferase domains, family N105 with remote homologies to potassium channel tetramerization domains, and families N73 and N430, which are similar to yeast and poxvirus transcription factors, respectively. Other interesting families that invite further investigation are N425 that contains the major capsid protein and the family pair N79 (transposase) / N80 (site-specific integrase-resolvase), which contains three adjacent pairs of transposase/resolvase genes (L79/R80, R104/L103, L770/R771).

**Discussion**

The here described ancient segmental duplications and massive ongoing individual gene duplications in Mimivirus are parsimonious with the postulated early evolutionary origins of this virus(21, 25, 29). It explains the origin of a large part of its genome without the need for over-proportional gene acquisition through horizontal gene transfer from a host organism, as this is commonly thought to be the case for smaller viruses. In fact, the gene duplication rate of Mimivirus lies with 38% well within the range of prevalence in the three domains of life, e.g. 17% for *Haemophilus influenza*, 44% for *Mycoplasma pneumoniae*, 30% for *Archaeoglobus fulgidus*, 30% for *Saccharomyces cerevisiae*, 38% for *Homo sapiens*, and 65% for *Arabidopsis thaliana* (36 and references therein). In this context, it is interesting to note that Ogata et al. (25) recently showed that horizontal gene transfer in Mimivirus is not more elevated than what is detected in bacteria neither.

Using multiple alignments together with remote homology detection methods based on Hidden Markov Model comparison, we attribute putative functions to some of the larger paralogous gene families. These attributions indicate that a number of these duplicated Mimivirus genes are likely to interfere with important host processes, such as transcription control, protein degradation, and different cell regulatory processes. The toleration and fixation of such important genome expansions under selective conditions may be explained by Mimivirus' particular life style, that is the fact that Mimivirus mimics a microbial prey to its amoebaean "predator" in order enter its host by phagocytosis. Thus, in order to represent an interesting prey for the amoeba, Mimivirus has to maintain bacterial size (15) and can thus tolerate more easily a large genome size than its smaller cousins. With this constraint comes the evolutionary advantage of being able to host a larger spectrum of genes capable to interfere with host defenses, very much in contrast to the situation of small viruses that are optimized for rapid and economic replication and that survive with a rather minimal gene set (for a detailed discussion see 5). Interestingly, if the same detection algorithm is applied to other large DNA viruses, a log-linear trend becomes visible between the number of paralogous genes and the gene content of the genome (Figure 5).

It is interesting to note that the larger families of proteins that are frequently repeated in tandem contain functions that are likely to play a role in virus-host interactions. Notable examples are the protein kinase family (N232), that may interfere with the host signaling network or other regulatory processes, the F-box containing cluster (N165), that may tag selected host proteins for destruction through the ubiquitin pathway, and the Zinc finger (BTB/POZ) family (N35) that may interfere with host transcription regulation. Unsurprisingly for such a large virus, two other large Mimivirus families of paralogues seem to play more structural roles, that is the largest family of all, the ankyrin repeat containing proteins and the collagen triple helix containing repeat proteins. The two families N172 (Lcluster) and N226 are particular intriguing, since no putative function could have been associated to these genes. The families are exceptionally well clustered, and underwent more recent duplications. It may therefore be speculated that they are related to more recent and novel function acquirements that may be specific to the lineage *Acantamoeba polyphaga Mimivirus*. Searching the Sargasso Sea environmental genome shotgun sequencing dataset (34), Ghedin and Claverie (10, submitted) detected the presence of close relatives of Mimivirus in this marine environment. While a large number of the Mimivirus genes are found to have a BLAST hit to this dataset, none of the genes from the N172 and N226 clusters (with the exception of a spurious match for gene L177) are found in the Sargasso Sea dataset. This may be an indicator for a more recent emergence of these two families.

The large fraction of viral genes that exhibit no or only remote homology to genes in any other organism, including different viruses (12), is commonly attributed to an assumed faster evolution of viral genes when compared to their bacterial and eukaryotic counterparts. If this assumption is correct, the genes of the two families N172 and N226 may have evolved from an ancient ancestor to a point where no similarity at the sequence level can be detected to their orthologues in other genomes.

Determining the 3-dimensional structure of members of these (and other) families may therefore answer the question as to the origin of these – at present Mimivirus-specific – genes. Comparing the structures of different paralogues may then contribute more generally to our understanding of the evolution of viral genes, as they have evolved in a unique environment in a singe genome context, i.e. in a situation where differences in G+C content or constrains related metabolic differences due to the availability of different amino acids are not to be considered.

We believe that gene and genome duplications in large DNA viruses can be analyzed much as it is presently done for members of the other three domains of life. For example, reconstructing duplication history has received extensive attention recently. Zhang *et al.* (38) present a method for inferring the duplication history of tandem repeated sequences that may be readily applied to Mimivirus tandem gene duplications. Davis and Petrov (6) demonstrated that genes that have generated duplicates in the *C. elegans* and *S. cerevisiae* genomes were 25%-50% more constrained prior to duplication than the genes that failed to leave duplicates. They further showed that conserved genes have been consistently prolific in generating duplicates for hundreds of millions of years in these two species, that is that the set of duplicate genes is biased. This observation may allow to narrow down the range of putative roles of the duplicated Mimivirus genes for which their function is still completely unknown.

Our analysis shows that a large fraction of the Mimivirus genes originates from repeated tandem gene duplications and from segmental genome duplication events, the order of magnitude of the duplications being comparable to what is commonly observed in bacteria, archeae and eukaryotes. This is compatible with the view that the large DNA viruses establish a deeply rooted branch on the tree of life rather than representing just a collection of genes, gathered during their passage in diverse cellular host organisms (see also discussion in 21, 25).

**Materials and Methods**

**Detection of paralogous genes** was performed using programs from the BLAST package (1). For the detection of paralogous families, each of the 911 Mimivirus genes was used to initiate a BLASTP search, followed by one or several PSI-Blast iterations until convergence. For the identification of homologous genes, all Mimivirus genes were compared to each other using BLASTP, where only the highest scoring match above a defined e-value cutoff was retained (best unidirectional match criteria). To test for a possible dependence on the choice of the e-value threshold, three different e-values $10^{-5}$, $10^{-10}$, $10^{-25}$ were used. If not stated otherwise, $10^{-10}$ is used as a reference e-value throughout this paper.

**Remote protein homology detection** was done by pair wise Hidden Markov Model (HMM) comparison using the HHsearch package (32) together with HMMs based on multiple alignments from the conserved domain database CDD (20), i.e. COG (33), SMART (17), PFAM (3), and SCOP (22). Multiple alignments of the paralogous genes were computed using the latest version of the T-Coffee package with advanced alignment options (23, 24, 28). Secondary structure predictions from PSIPRED (13) were included in the HMM-HMM comparison as described in (32). Results of the HMM search and multiple alignments are available at http://igs-server.cnrs-mrs.fr/suhre/mimiparalogues/

**Genome sequences** of all fully sequenced viral genomes (as of Nov. 2004) were downloaded from the National center for biotechnology information (NCBI) viral genomes project (2) at http://www.ncbi.nlm.nih.gov/genomes/VIRUSES/viruses.html. All (223) genomes with more than 50 annotated genes were included in the analysis.

**Acknowledgments**

This work has been supported by CNRS and Marseille-Nice Génopole. I wish to thank Johannes Söding for assistance with the use of the *HHsearch* program and acknowledge helpful discussions with my colleagues at the laboratory IGS, in particular C. Abergel, S. Audic, G. Blanc, C. Notredame, H. Ogata, and J.-M. Claverie.

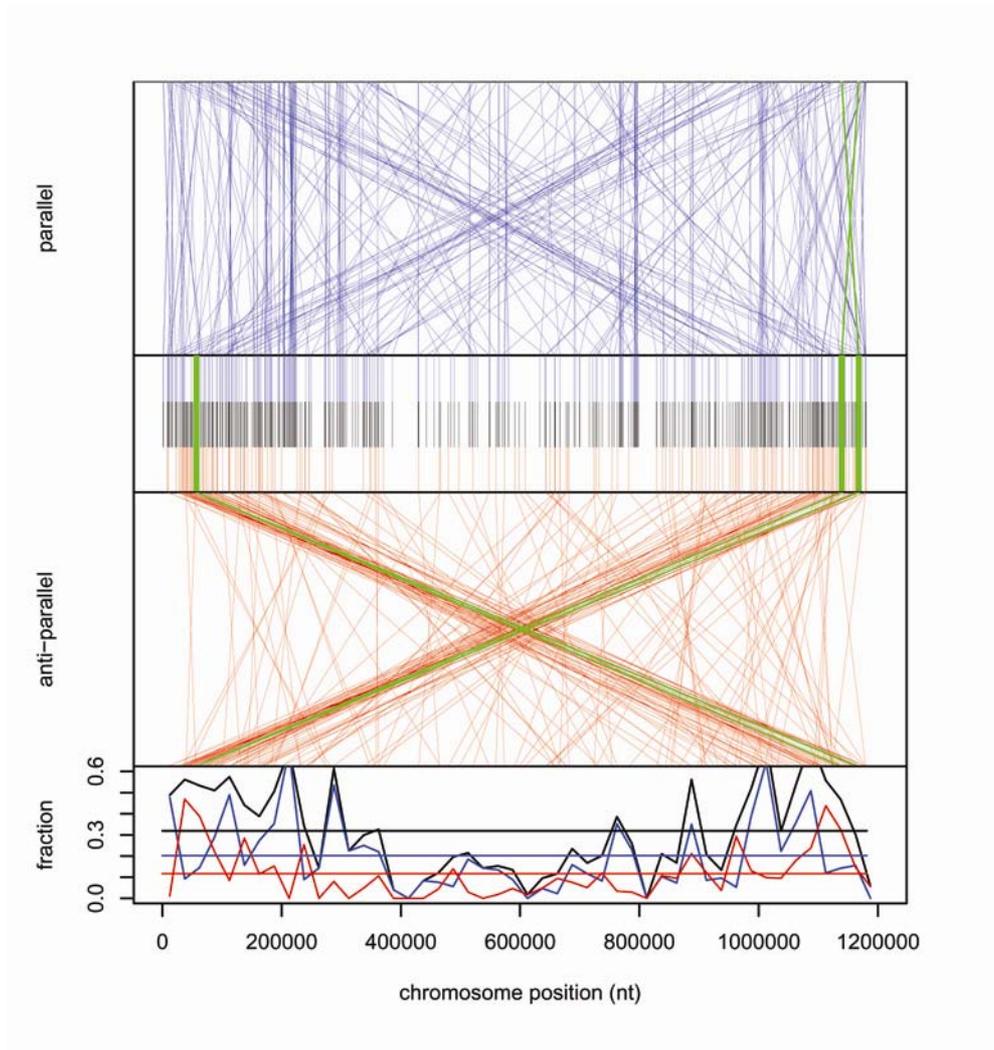

**Figure 1**

Correspondence between homologous genes in the Mimivirus genome; shown are the best BLASTP hits for all 911 Mimivirus genes (e-value = $10^{-10}$); blue lines indicate gene pairs that are duplicated in parallel orientation (*cis*), that is genes that are either found both on the strand coding in the positive or both on the strand coding in the negative direction; red lines correspond to genes duplicated in anti-parallel orientation (*trans*); the top frame and the 3$^{rd}$ frame from the top connect homologous genes represented as their position on the chromosome, the 2$^{nd}$ frame from the top indicates the chromosomal location of all duplicated genes; the bottom frame gives the chromosomal fraction that is duplicated in parallel (blue), in anti-parallel direction (red), and both (black) averaged over a window size of 25,000 nucleotides. Green lines indicate the position of the three tRNA-Leu genes that were identified in the Mimivirus genome.

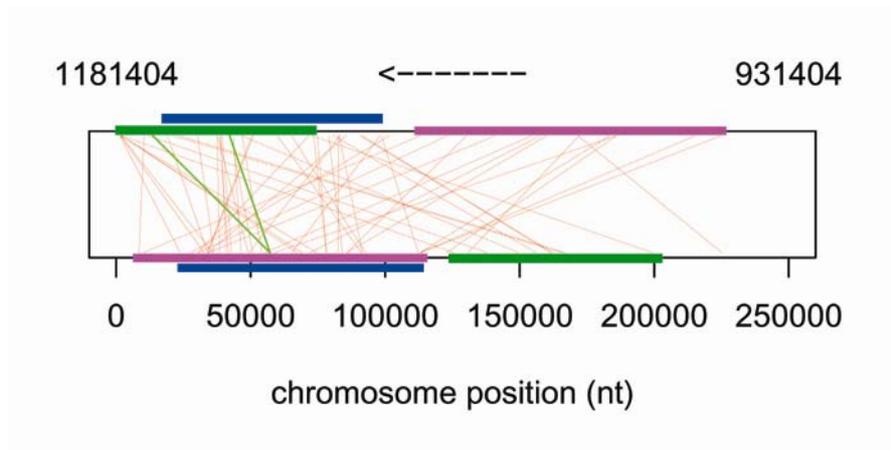

**Figure 2**

Correspondence between anti-parallel duplicated genes (*cis*) between the two "telomeric" regions, 0–250,000 and 931,404–1,181,404; the latter region is presented in reversed direction. Potentially syntenic regions are marked by identical colors.

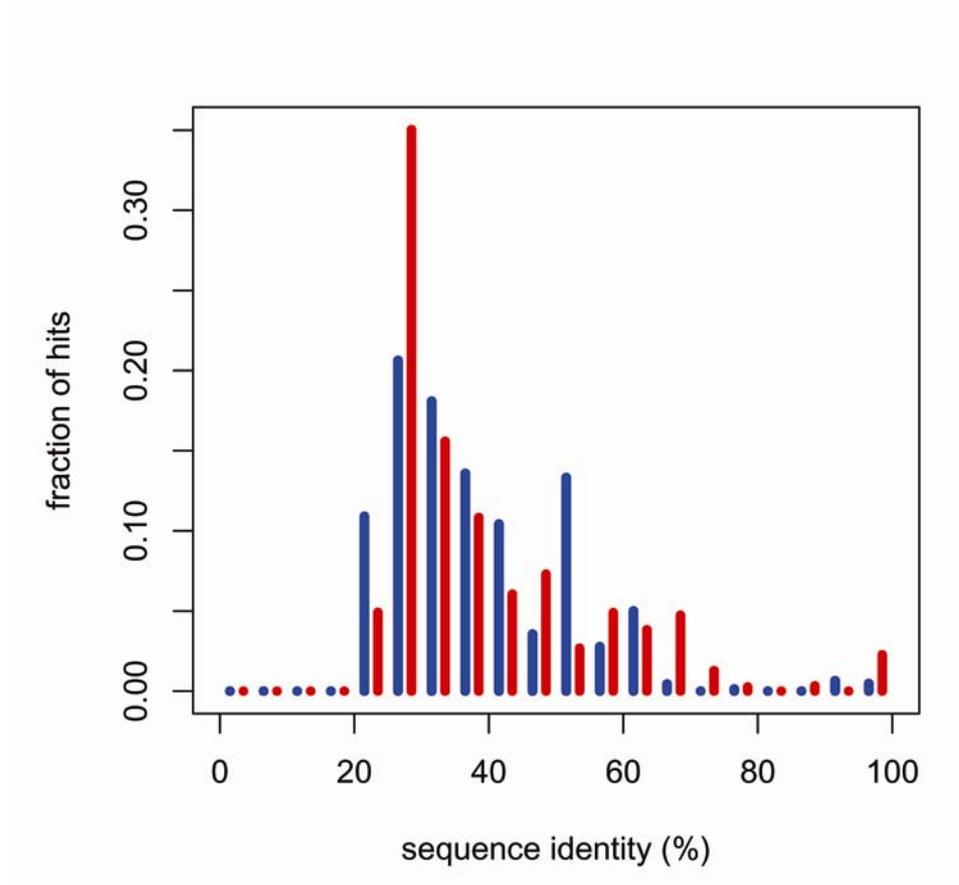

**Figure 3**

Distribution of BLASTP hits (weighted by the alignment length) as a function of sequence identity between matching genes (interval size is 5%); blue: genes duplicated in parallel direction; red: genes duplicated in anti-parallel direction.

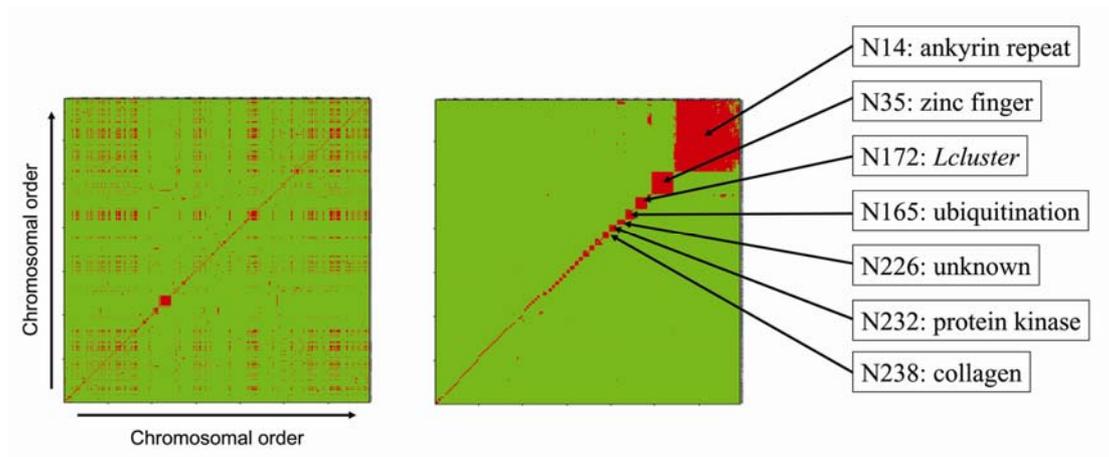

**Figure 4**

Dotplot of paralogous genes as a function of their position on the Mimivirus chromosome (left) and clustered by paralogous family (right). Matching genes are marked by red dots. Paralogous families are named by the gene that was used to initiate the PSI-BLAST search (e.g. the N14 family was seeded using gene L14). High resolution images with annotated gene names are provided as supplementary material.

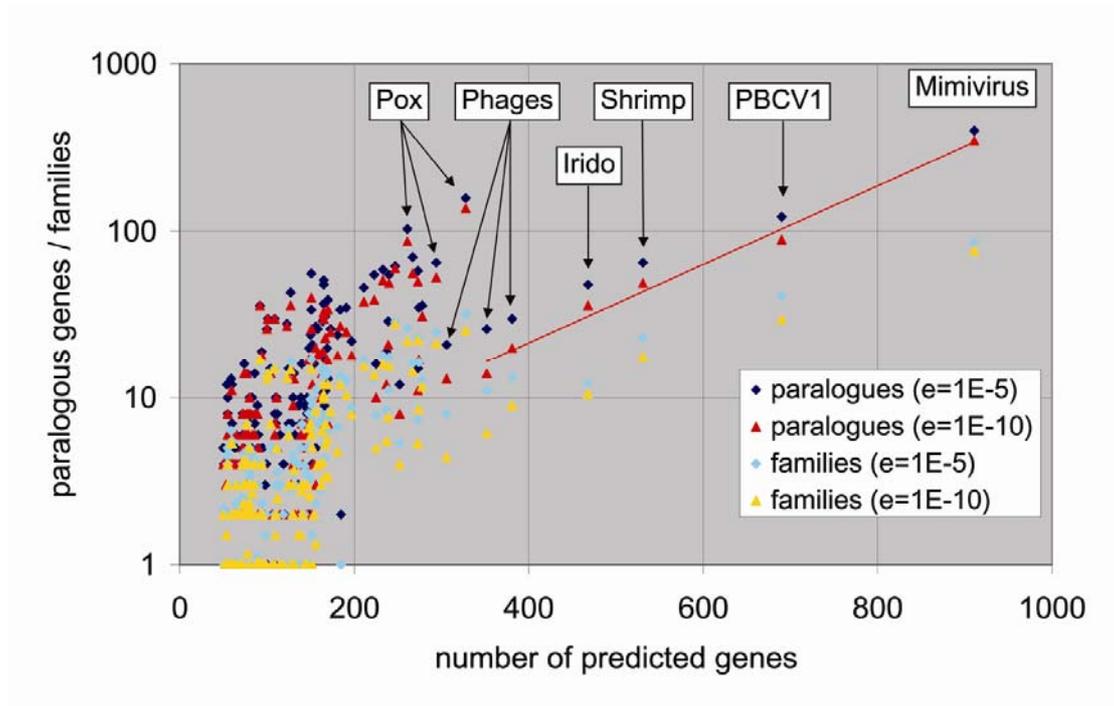

**Figure 5**

Number of paralogous genes and paralogous gene families as a function of the number of predicted genes in the genomes. Results are presented for two different e-values, $10^{-5}$ (blue/cyan) and $10^{-10}$ (red/orange). Abbreviations are "PBCV1" (*Paramecium bursaria Chlorella virus 1*), "Shrimp" (*Shrimp white spot syndrome virus*), "Irido" (*Invertebrate iridescent virus 6*), "Phages" - from largest to smallest (*Bacteriophage KVP40/ Bacteriophage Aeh1/ Pseudomonas phage phiKZ*), "Pox" - from largest to smallest (*Canarypox virus / Amsacta moorei entomopoxvirus / Fowlpox virus*). For the large DNA viruses, a log-linear relationship between gene duplication and gene content is observed (red line).

**Table 1**

Gene names and annotation of the members of the largest paralogous families. Genes are numbered in increasing order by their position on the linear Mimivirus chromosome. The letter 'L' indicates genes that are transcribed 'to the left' (negative strand), the letter 'R' stands for genes transcribed 'to the right' (positive strand). Tandem (*Cis*) duplications can be identified by successive numbering and identical letters (e.g. L121 and L122 are adjacent genes that are coded on the same strand). A full list with all families is available as supplementary material.

N14 – Ankyrin repeat :
- L14 L22 L23 L25 L36 L42 L45 L56 L59 L62 L63 L66 L72 L88 L91 L93 L99 L100 L109 L112 L120 L121 L122 L148 R229 R267 L279 L482 L483 R579 L589 R600 R601 R602 R603 R634 L675 L715 R760 R777 R784 R787 R789 R791 R797 R810 R825 R835 R837 R838 R840 R844 R845 R846 R847 R848 L863 L864 R873 R875 R880 R886 R896 R901 R903 R911

N35 – Zinc finger (BTB/POZ, some with WD-repeats) :
- L35 L49 L55 R61 L67 L76 L85 L89 L98 L107 R154 R224 R225 L272 L344 R731 R738 R739 R765 R773 L783 L786 L788 R830 L834 R842

N172 – Function unknown (*Lcluster*) :
- L172 L174 L175 L176 L177 L178 L179 L180 L181 L182 L183 L184 L185 L697

N165 – Ubiquitination related (N-terminal F-box, FNIP-repeats) :
- L60 L162 L165 L166 L167 L168 L170 R286 L414 L415

N226 – Function unknown :
- L226 L228 R734 L764 L766 L767 L768 L769 L774

N232 – Protein kinase domain :
- L232 L268 R436 R517 L670 L673 R818 R826 R831

N238 – Collagen triple helix repeat containing proteins :
- L71 R196 R238 R240 R241 L668 L669